\documentclass[prb,superscriptaddress,twocolumn,10pt,aps]{revtex4-1}
\usepackage{epsfig}
\usepackage{amssymb,amsmath}
\usepackage{graphicx}
\usepackage{amsfonts}
\usepackage{epstopdf}
\usepackage{tikz}
\usepackage{hyperref}
\hypersetup{colorlinks=true,citecolor=blue,linkcolor=blue,filecolor=blue,urlcolor=blue,pdfpagemode=UseNone}

\bibpunct{[}{]}{,}{n}{}{}   
\begin{document}

\title{Investigating transverse Hall conductances using two-terminal setups}

\author{Alexis R. Hern\'andez}
\affiliation{Instituto de F\'{\i}sica, Universidade Federal do Rio de Janeiro, 
21941-972 Rio de Janeiro, Brazil}

\author{Leandro R. F. Lima}
\affiliation{Instituto de F\'{\i}sica, Universidade Federal Fluminense, 
24210-346 Niter\'oi, Brazil}


\date{\today}

\begin{abstract}
In this paper we present a method to numerically study transverse Hall conductances using a two-terminal setup.
Using nonlinear transport concepts
we find that the Hall voltage dependence on the model parameters can be investigated from the difference between the injectivities from each terminal. 
The method is suitable to work with non-equilibrium Green's functions as well as for scattering matrix approaches. 
We illustrate the proposed idea by studying the quantum spin Hall effect in graphene with disordered spin-orbit scattering
centers induced by adatoms.
We use two distinct models, 
a finite difference implementation of the Dirac Hamiltonian and a tight-binding Hamiltonian combined with the scattering matrix approach and the non-equilibrium Green's functions approach, respectively.  
\end{abstract}


\maketitle

\section{Introduction}
\label{sec:introduction}

The bloom of topological ideas in the condensed matter community is one of the driving forces which propel new discoveries in the field. 
The most prominent topological effect of quantum matter is the quantum Hall effect (QHE) \cite{Wang17,vonKlitzing80}, where a strong magnetic field, perpendicular to the sample, leads to an electronic current that flows only through the sample edges and it is robust against backscattering.

In the last two decades, another topological effect, called quantum spin Hall effect (QSHE) \cite{Wang17}, was theoretically predicted \cite{Bernevig06} and experimentally confirmed \cite{Konig07}.
Here the spin-orbit coupling (SOC) gives rise to spin polarized charge propagation through opposite edges of the sample \cite{Wang17}. 
The QSHE has been at the spotlight ever since its observation on HgTe/CdTe quantum wells \cite{Konig07}.
The latter has triggered intense investigation on how the robustness of the spin-polarized current against non-magnetic disorder and how it is affected by time reversal symmetry breaking caused by an external magnetic field \cite{Maciejko10,Scharf12,Durnev16,Nanclares17}.

There are other topological states associated with edge/surface current propagation in condensed matter physics such as the states in the quantum anomalous Hall effect, in chiral topological superconductors and in Weyl semimetals \cite{Wang17}.    
The protagonist in almost any case are the chiral states which arise as a consequence of non-trivial topologies. 
To address these states, experimentalists usually need samples with four- or six-terminal geometries to measure the transverse charge or spin conductances.
However, the theoretical study of such systems with realistic sizes and disorder demands the use of numerical methods, such as the recursive Green's function method \cite{MacKinnon85,Lewenkopf13}.
Although methods to treat the electronic transport in multi-terminal setups have been developed along the years \cite{Baranger88,Kazymyrenko08,Thorgilsson14,Lima18}, the use of multiple terminals undermines the size and speed capabilities of the method as compared to a simple two-terminal setup.

In this paper we describe how to surround this shortcoming by alternatively using two-terminal calculations to study transverse conductivities. 
To illustrate the method we address the QSHE in graphene doped with adatoms \cite{Weeks11,Balakrishnan14} using two different methods: A finite difference one, which computes the scattering matrix for Dirac particles \cite{DFGAlexis12} in a strip geometry and the usual tight-binding description \cite{CastroNeto09} combined with non-equilibrium Green's functions.

The paper is organized as follows: 
Section~\ref{sec:nonlinear} introduces the general ideas from non-linear transport that are relevant to this study. 
Next we describe how to study transverse conductances using those concepts in Sec.~\ref{sec:hall}.
We devote Sec.~\ref{sec:QSHE} to illustrate the method by investigating the spin accumulation at the edges of graphene nanoribbons due to the presence of disordered spin-orbit scattering centers.
We conclude in Sec.~\ref{sec:conclusions}.

\section{Nonlinear transport}
\label{sec:nonlinear}


The Landauer-Buttiker formula is a cornerstone in the study of electronic transport in mesoscopic systems. 
It allows for the computation of the electronic current, at a particular contact, in terms of the electronic transmission probabilities. Following the notation of Ref.~\onlinecite{Hernandez13}, the Landauer-Buttiker formula for the electronic current at terminal $\alpha$ writes
\begin{align}
	\label{eq:LB}
	I_{\alpha}=\frac{2 e}{h}\sum_{\beta=1}^N \int_{-\infty}^{\infty} \!
	dE\, f_\beta(E)\, A_{\alpha\beta}[E,U(\mathbf r)] \;,
\end{align}
{
were $f_\beta(E) = f_0(E-eV_{\beta})$ and $V_\beta$ are the electronic distribution and potential at the contact $\beta$, respectively, $f_0(E)=(e^{E/k_{\rm B}T} + 1)^{-1}$ is the equilibrium Fermi-Dirac distribution function at temperature $T$ and the factor $2$ accounts for the spin degeneracy. The quantity $A_{\alpha\beta}[E,U(\mathbf r)]$, 
}
that encodes the transmission properties of the system, is expressed in terms of the scattering matrix $\mathbf S_{\alpha\beta}$ as
\begin{align}
	A_{\alpha\beta}[E,U({\bf r})]
	= {\rm Tr} [{\bf 1}_{\alpha}\delta_{\alpha\beta} - {\bf S}^\dagger_{\alpha\beta}{\bf S}^{}_{\alpha\beta}].
	\label{eq:A}
\end{align}
In Eqs.~(\ref{eq:LB}) and (\ref{eq:A}) we make explicit the dependence of the transmission amplitudes on the electrostatic potential inside the system $U({\bf r})$. 
In linear response, $A_{\alpha\beta}$ is computed at
the equilibrium potential $U_{\rm eq}({\bf r})$ which is established when all reservoirs
have the same equilibrium chemical potential $\mu_0$. Beyond this regime, it is necessary to compute
$U({\bf r})$ self-consistently, as pointed out by Landauer \cite{Landauer87}.

To make analytical progress, it is convenient to expand all quantities in powers of $V_\alpha$.
The local electrostatic potential $U({\bf r})$ reads
\begin{align}
\label{eq:U}
U({\bf r})= U_{\rm eq}({\bf r}) + \sum_{\alpha}u_{\alpha}({\bf r})V_{\alpha} + O(V_\alpha^2)
\end{align}
where  $u_{\alpha}({\bf r})$ is the characteristic potential defined by
\begin{align}
\label{eq:defu_char}
u_{\alpha}({\bf r})= \frac{\partial}{\partial V_{\alpha}} U({\bf r}) \Big|_{\{V_\gamma\}=0}.
\end{align}
Here $\{V_\gamma\}=0$ is a shorthand for $V_\gamma=0$, for all $\gamma$.

To determine $u_{\alpha}({\bf r})$, we need a self-consistent microscopic electronic structure calculation, or an adequate approximation.
The latter was developed in Ref.~\onlinecite{Buttiker93} assuming that the potential $U({\bf r})$ is related to the electronic density imbalance $\delta n({\bf r})$ generated by the bias. 
In turn, $\delta n({\bf r})$ arises from the charge injected by the leads, $dn_{inj}({\bf r})$, and the induced charge in the conductor due to the injected one, $dn_{ind}({\bf r})$.

At linear order, the injected charge $dn_{inj}({\bf r})$ is proportional to the injection properties of the sample which is given by the injectivity, namely
\begin{align}
\label{eq:inj}
\frac{dn(\mathbf r,\alpha )}{dE} =
&-\frac{1}{2\pi i}\int_{-\infty}^{\infty} dE \left( -\frac{\partial f_0}{\partial E}\right)
\nonumber\\
&\times \sum_{\beta}{\rm Tr} \left[ {\bf S}^{\dagger}_{\beta \alpha}\frac{\delta {\bf S}_{\beta \alpha}}
{e\delta U({\bf r})}-\frac{\delta {\bf S}^{\dagger}_{\beta\alpha}}{e\delta U(\mathbf r)}
{\bf S}_{\beta\alpha}\right],
\end{align}
evaluated at $\{V_\gamma\}=0$. Here we included the factor two due to spin degeneracy.
The injectivity describes the linear contribution to the local density of states related to incoming states from a given contact.

The induced charge density, in linear order of $V$, is given by
\begin{align}
dn_{\rm ind}({\bf r}) = e\sum_\alpha   \int d{\bf r}^\prime \,
\Pi({\bf r},{\bf r}^\prime)\, u_\alpha({\bf r}^\prime) dV_\alpha,
\end{align}
where $\Pi ({\bf r},{\bf r}')$ is the Lindhard polarization function \cite{Bruus-Flensberg04}. 
The scattering approach does not provide a recipe to obtain the latter. However, by recalling the relation between the Wigner-Smith time delay and the conductor density of states, $dn_{\rm ind}({\bf r})$ can be readily written in the Thomas-Fermi approximation as \cite{Buttiker93}
\begin{align}
\label{eq:ThomasFermi}
dn_{\rm ind}({\bf r}) = e\sum_\alpha  \frac{dn({\bf r} )}{dE}\, u_\alpha({\bf r}) dV_\alpha .
\end{align}
The local density of states $dn/dE$ is
\begin{align}
\frac{dn({\bf r} )}{dE}=
\sum_{\beta} \frac{dn(\beta, {\bf r} )}{dE},
\end{align}
where $dn(\beta, {\bf r} )/dE$ is called emissivity and is given by
\begin{align}
\frac{dn(\beta, {\bf r} )}{dE} =
&-\frac{1}{2\pi i}\int dE \left( -\frac{\partial f_0}{\partial E}\right)
\nonumber\\
&\times \sum_{\alpha} {\rm Tr}\left[ {\bf S}^{\dagger}_{\beta \alpha}\frac{\delta {\bf S}_{\beta \alpha}}{e\delta U({\bf r})}-\frac{\delta {\bf S}^{\dagger}_{\beta\alpha}}{e\delta U({\bf r})}{\bf S}_{\beta\alpha}\right].
\label{eq:emi}
\end{align}
Analogously, the emissivity describes the linear contribution to the local density of states related to outgoing states throughout a given contact.

These elements render the Poisson equation
\begin{align}
\label{eq:scatPoisson}
-\nabla^2 u_{\alpha}({\bf r})+4\pi e^2 \frac{dn({\bf r} )}{dE}u_{\alpha}({\bf r})=4\pi e^2  \frac{dn({\bf r},\alpha )}{dE},
\end{align}
where both the density of states and the injectivity depend only on the scattering matrix.

\section{Charge and Spin Hall Conductances}
\label{sec:hall}

\begin{figure}[!t]
\includegraphics[width=1\columnwidth]{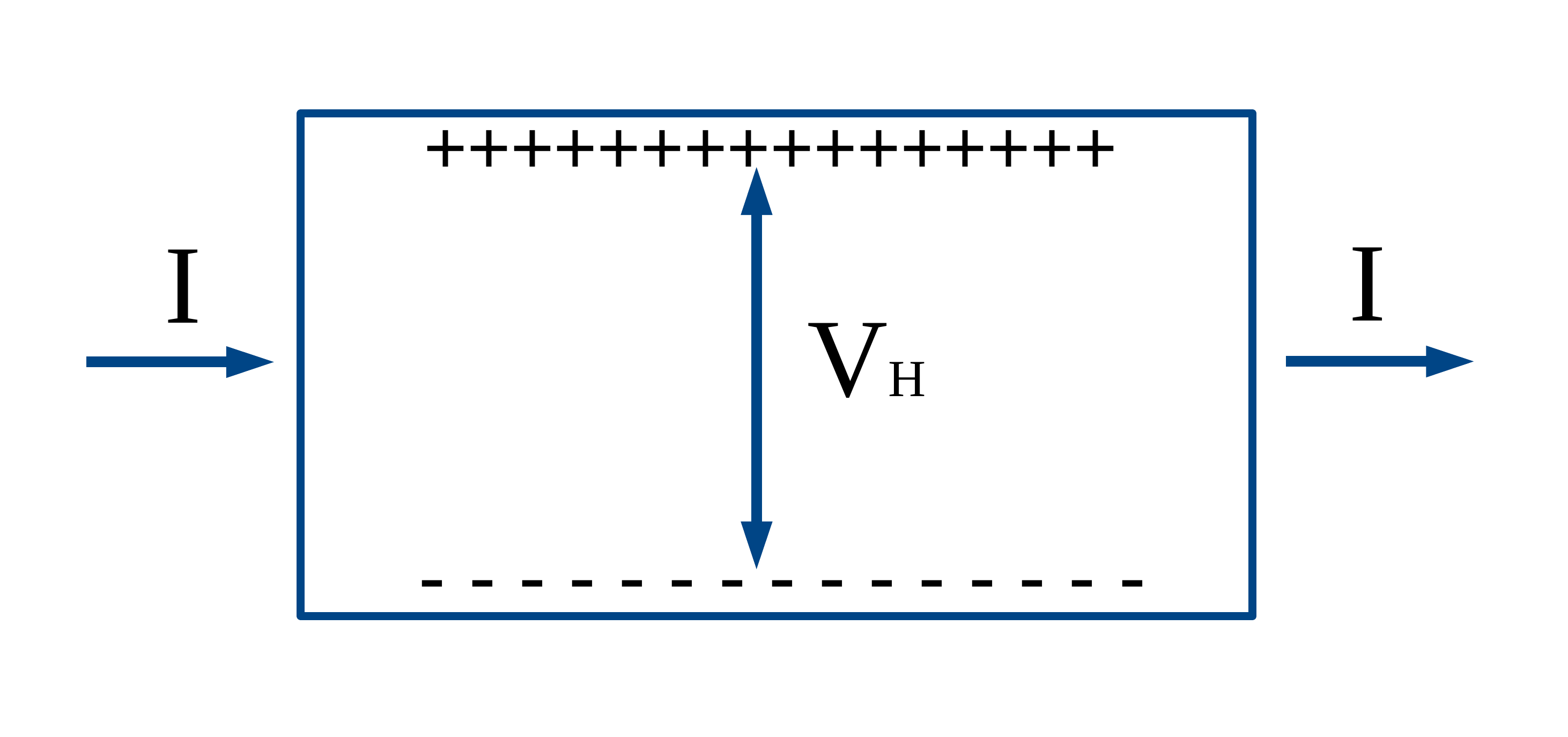}
\caption[]{Schematic view of the system. The electronic current $I$ enters through the terminal $1$ (left) and leaves through the terminal $2$ (right). The upper (lower) egde is positively (negatively) charged giving rise the a Hall voltage $V_H$. The magnetic field is perpendicular to the system. }\label{fig:HB}
\end{figure}

In the presence of a perpendicular magnetic field, the current $I$ passing through the system illustrated in Fig.~\ref{fig:HB} generates a Hall voltage $V_H$ established in the transverse direction due to charge accumulation at the edges.
The Hall conductance in such a system is defined as
\begin{align}
\sigma_H = \frac{I}{V_H}.
\end{align}

For a two-terminal system as the one in Fig.~\ref{fig:HB}, the electronic current $I$ is given by the Landauer Formula in Eq.~(\ref{eq:LB}).
For a small bias, we evaluate $I$ at $U_{eq}({\bf r})$ and compute the Hall voltage $V_H$ in terms of the characteristic potentials $u_\alpha$ by means of Eq.~(\ref{eq:U}). 
In linear response, $V_H$ reads
\begin{align}
V_H &= U(r_+)-U(r_-),\\ 
&= [u_1(r_+)V_1+u_2(r_+)V_2] -[u_1(r_-)V_1+u_2(r_-)V_2] \nonumber 
\end{align}
where $r_+$ and $r_-$ correspond to the edges of the system. 
For a symmetric applied voltage, \textit{i.e.} $V_1=-V_2=V/2$, we find
\begin{align}
V_H &= \left[ \Delta u(r_+) - \Delta u(r_-) \right] V/2,
\label{hallvoltage}
\end{align}
where $\Delta u(r) \equiv u_1(r)-u_2(r)$.

The Hall voltage $V_H$ in Eq.~(\ref{hallvoltage}) depends on the two-terminal characteristic potential difference $\Delta u(\rm r)$ instead of the individual values $u_\alpha$. 
From Eq.~(\ref{eq:scatPoisson}) we find that $\Delta u(\rm r)$ is given by 
\begin{align}
	\label{eq:scatPoissonu}
	-\nabla^2 \Delta u(\mathbf r)+4\pi e^2 \frac{dn(\mathbf r)}{dE}\Delta u(\mathbf r)=4\pi e^2  \Delta n(\mathbf r),
\end{align}
where $\Delta n(\rm r) \equiv \frac{dn({\bf r},1)}{dE} - \frac{dn({\bf r},2)}{dE}$ is the two-terminal injectivity difference at the position $\mathbf r$.
Thus, the Hall voltage $V_H$ is a function of $\Delta n(\mathbf r_{\pm})$ as well, namely
 \begin{align}
	V_H &= \left\{ \Delta u[\Delta n(r_+)] - \Delta u[\Delta n(r_-)] \right\} V/2.
	\label{injdiff}
 \end{align}

Equation~(\ref{injdiff}) states that the QHE is manifested only if the charge injection density imbalance $\Delta_{\pm} \equiv \Delta n(r_+)- \Delta n(r_-)$ is nonzero and that $V_H$ increases with the imbalance $\bar\Delta n$. 
The latter is true since $\Delta u$ monotonically increases with $\Delta n$, see Eq.~(\ref{eq:scatPoissonu}), so that a larger imbalance $\Delta n(r_+) - \Delta n(r_-)$ renders a larger difference $\Delta u(r_+) - \Delta u(r_-)$, which is proportional to $V_H$ for the case of symmetric bias in Eq.~(\ref{injdiff}).

Although the analytical dependence of the Hall voltage $V_H$ on $\Delta n$ in Eq.~(\ref{injdiff}) is unknown, it is possible to study the injection density $\Delta n$, or its imbalance $\Delta_{\pm}$, in order to determine (i) whether the QHE is present ($\Delta_{\pm} \neq 0 \Rightarrow V_H \neq 0$) or not ($\Delta_{\pm} = 0 \Rightarrow V_H = 0$) and (ii) if the QHE becomes stronger or weaker by varying any model parameter, since $V_H$ varies monotonically with $\bar\Delta n$.
This is the central result of this paper.

Spin dependent effects, such as the QSHE, cannot be directly quantified by the characteristic potentials in Eq.~(\ref{eq:scatPoisson}) because they do not distinguish the spin degrees of freedom.
On the other hand, one can calculate the spin resolved injectivities $\frac{dn^s(\mathbf r,\alpha)}{dE}$, where $s=\uparrow,\downarrow$ labels the spin, to obtain the two-terminal spin-resolved injection densities
\begin{align}
		\Delta n^s(\mathbf r) &= \left(\frac{dn^s(\rm r,1)}{dE} - \frac{dn^s(\rm r,2)}{dE}\right).
		\label{eq:dns}
\end{align}
Thus, it is straightforward to extend the ideas discussed for the Hall voltage $V_H$ to study the spin Hall voltage $V_{sH}$ by studying $\Delta_{\pm}^s \equiv \Delta n^s(r_+) - \Delta n^s(r_-)$, where $\Delta n^s$ is given by Eq.~(\ref{eq:dns}).
In this case, the spin-resolved imbalance $\Delta_{\pm}^s$ generates the spin Hall Voltage $V_{sH}^s$ for the spin orientation $s=\uparrow,\downarrow$.

In the next sections we illustrate the use Eq.~(\ref{eq:dns}) to quantitatively analyze the dependence of the QSHE in graphene nanoribbons doped with spin-orbit scatterers on a few model parameters. 

\section{Quantum Spin Hall effect: Graphene nanoribbons with disordered spin-orbit coupling}
\label{sec:QSHE}

We study local disordered spin-orbit coupling on graphene due to the presence of adatoms deposited on top of the graphene sheet. 
This system has been shown to present the QSHE both theoretically \cite{AdatomsSOC,Weeks11} and experimentally \cite{Balakrishnan14}. 
First we study numerically the density of spin accumulation for a finite size graphene nanoribbon doped with adatoms using the scattering matrix approach applied to the effective low energy continuous description given by the Dirac Hamiltonian.
Then we analyze the same system described by a full tight-binding Hamiltonian for graphene with effective local hoppings that mimick the presence of adatoms.

\subsection{Scattering matrix approach}
\label{sec:smatrix}

In this section we use the finite difference method presented in Ref.~\onlinecite{DFGAlexis12} to compute the scattering matrix of massless Dirac particles with spin orbit coupling disorder. The Hamiltonian is
\begin{equation}
	\label{eq:H_Dirac}
	H=-i\hbar v(\sigma_x\partial_x+\sigma_y\partial_y) + U({\bf r}) + U_{AD}({\bf r}) 
\end{equation}
where $v$ is the velocity of the massless Dirac fermions, $\sigma_x$ and $\sigma_y$ are Pauli matrices, $U({\bf r})$ is the electrostatic potential and  $U_{AD}(\mathbf r)=U_{SO}+U_{C}$ is the spin-orbit term that appear due to the presence of adatoms.
The adatoms are randomly distributed in the system with a concentration $n_{SOC}$.
The first contribution to the spin-orbit potential is given by $U_{SO}(\mathbf r) = \sum_j U_{SOC}\,\sigma_z \times \tau_z\, \delta ({\mathbf{r-\tilde{r}_j}}) $, where the constant $U_{SOC}$ is the spin-orbit strength, ${\mathbf{ \tilde{r}_j}}$ indicates the position where adatoms are located and the Pauli matrices $\sigma_z$ and $\tau_z$ refer to the pseudo-spin (sublattice) and spin degrees of freedom, respectively.
The adatoms also contribute with a on-site coulomb perturbation $U_{C}(\mathbf r) = \sum_j U_{C} \delta ({\mathbf{ r-\tilde{r}_j}})$.
The Hamiltonian includes a purely coulomb contribution $U(\mathbf r) = \sum_l U_{Dis} \delta ({\mathbf{ r-\bar{r}_l}})$ due to non-magnetic impurities, where $U_{Dis}$ is the impurity disorder strength and ${\mathbf{\bar{r}_j}}$ indicates the inpurities positions.
These impurities are also randomly distributed with a concentration $n_{Dis}$.

\begin{figure}[!t]
\includegraphics[width=0.9\columnwidth]{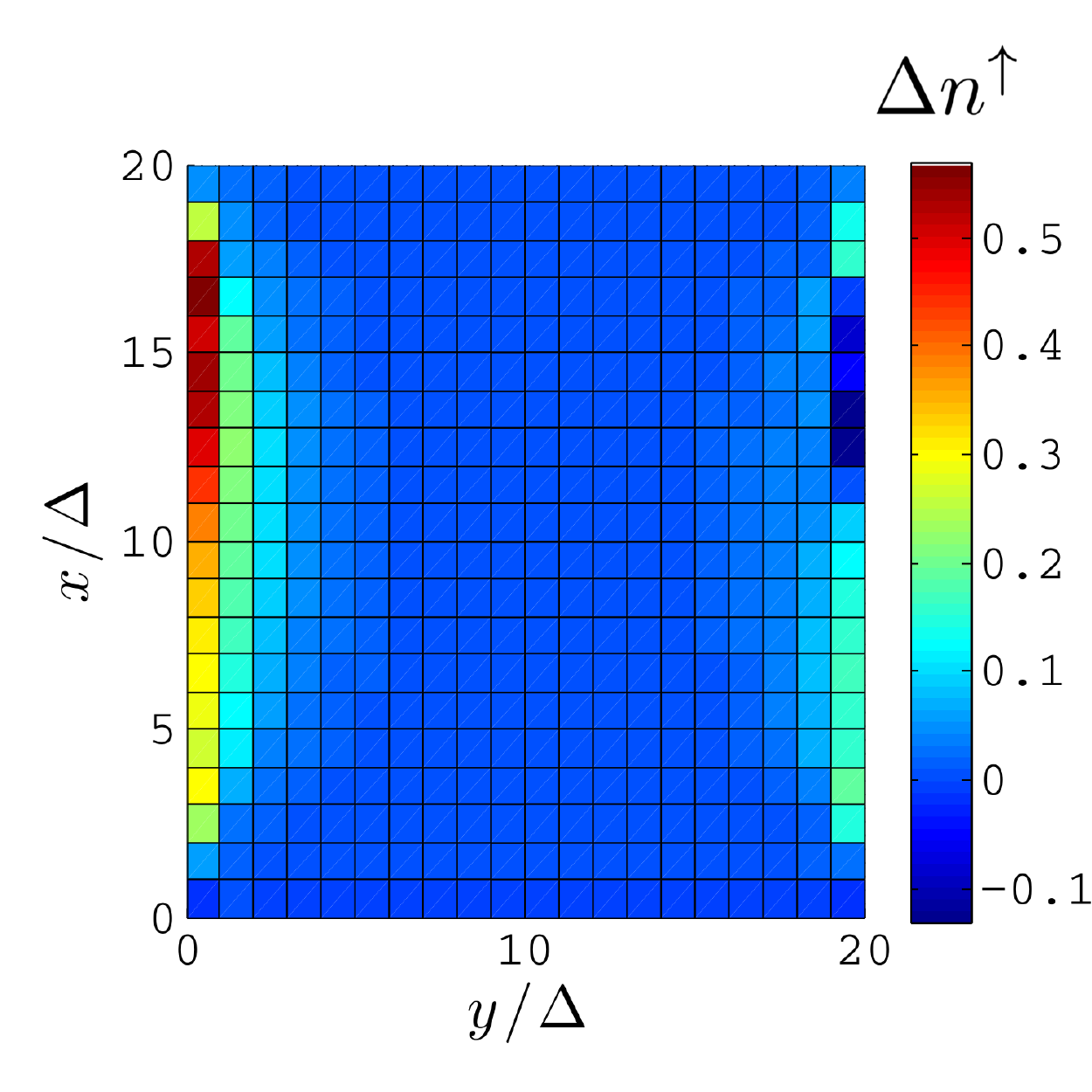}
\caption{
Injection density of spin up electrons $\Delta n^\uparrow$ in units of $1/\Delta^2 V$ for a single realization. The system has width $W=20\Delta$ and length $L=20\Delta$, where $\Delta= 5a_0$, $a_0=0.142\ nm$. The disorder concentration is $5 \%$, the SOC disorder strength is $U_{SOC}=0.02\ eV$ and the electronic energy is $E=-0.001\ eV$.
}
\label{fig:Injvsx}
\end{figure}

In order to study the QSHE we need to compute the injectivities for this system. 
To do that we numerically calculate the functional derivative of the scattering matrix as
\begin{align}
	\frac{\delta \mathbf S_{\alpha \beta}}{\delta U({\bf r_0})}=\lim_{\xi \rightarrow 0}
	\frac{\mathbf S_{\alpha\beta}\left[U({\bf r}+\xi \delta({\bf r}-{\bf r_0}))\right]-\mathbf S_{\alpha\beta}\left[U({\bf r})\right]}{\xi},
\end{align}
substitute the result in Eq.~(\ref{eq:inj}) and evaluate the injectivities at zero temperature.
We discretize the system using the lattice spacing $\Delta= 5a_0$, where $a_0=0.142nm$.
The nanoribbon has width $W=20\Delta$ and length $L=20\Delta$, see Fig.~\ref{fig:Injvsx}.
The system is attached to vertical semi-infinite leads at the positions $x=0$ and $x=L$.

\begin{figure}[!t]
\begin{tabular}{cc}
\includegraphics[width=0.49\columnwidth]{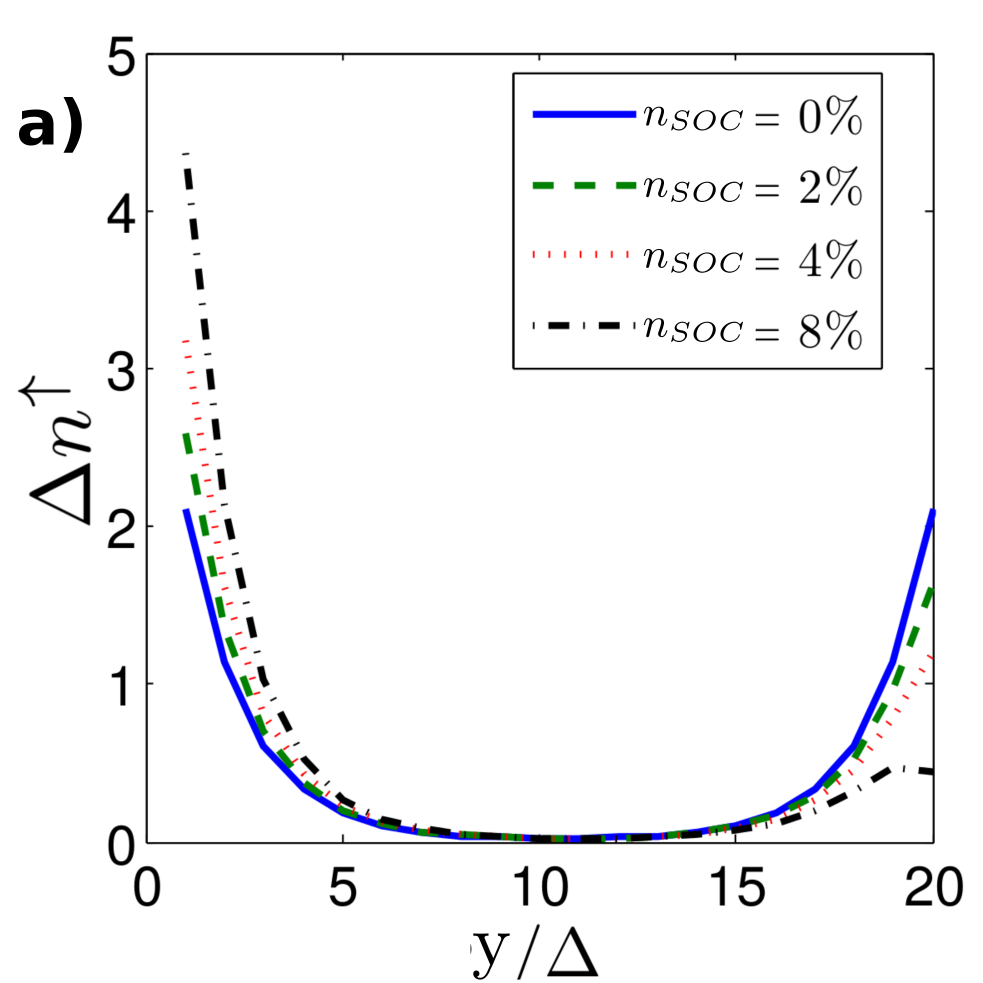} & 
\includegraphics[width=0.49\columnwidth]{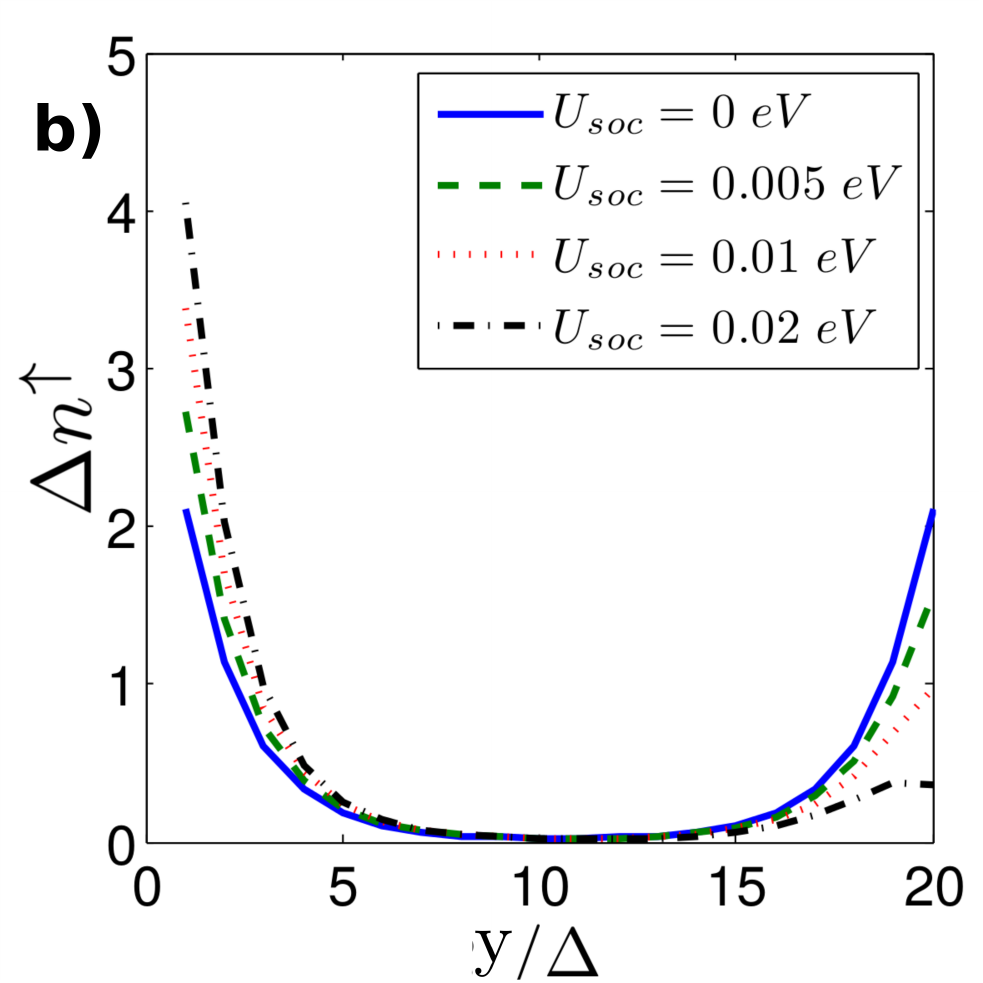} \\
\includegraphics[width=0.49\columnwidth]{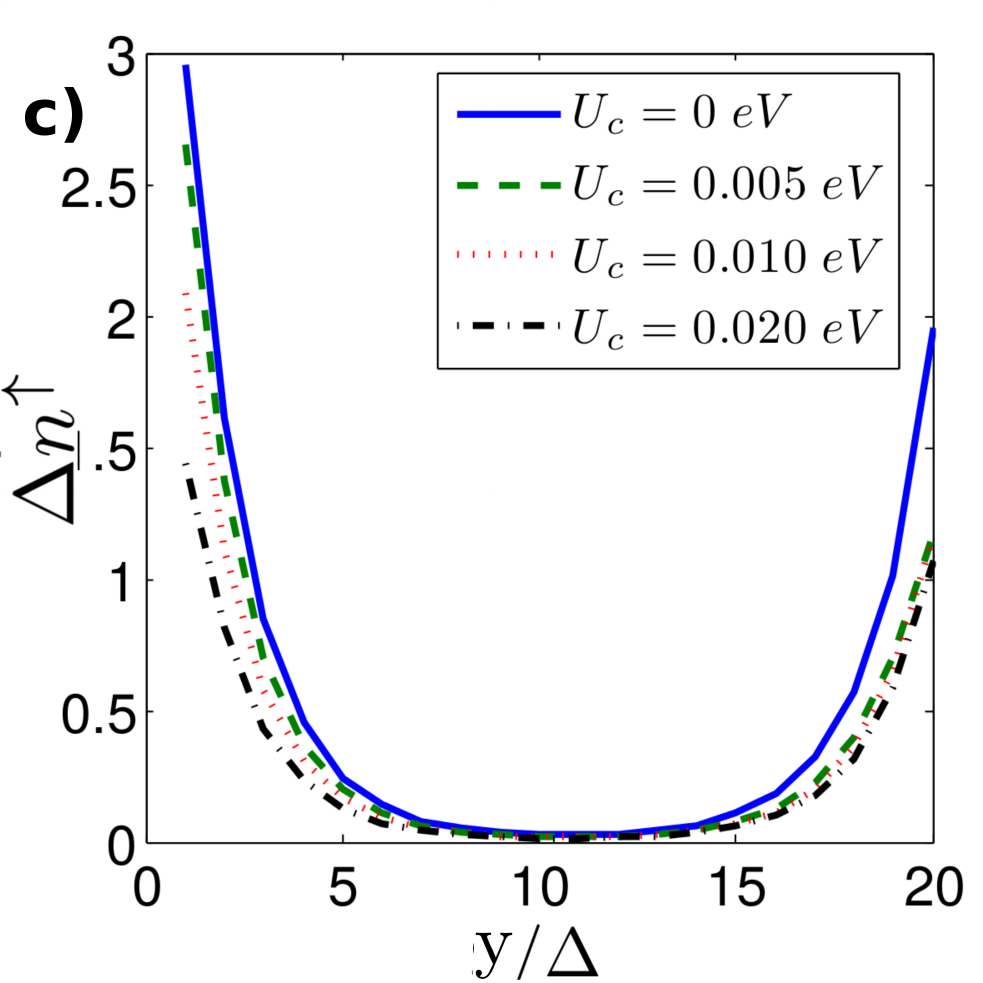} &
\includegraphics[width=0.49\columnwidth]{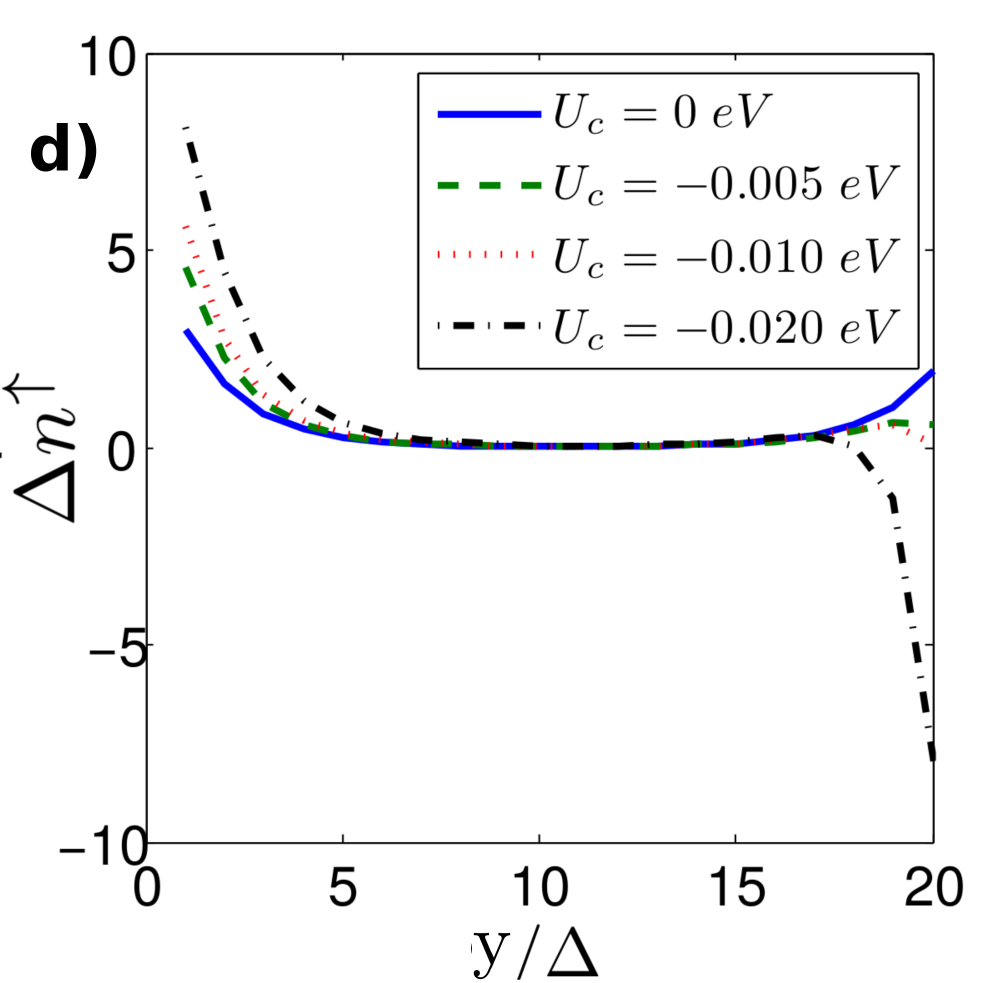}  
\end{tabular}
\caption{
Spin-up injection density $\Delta n^\uparrow$ in units of $1/(\Delta^2 V)$ calculated at $x=L/2$ as a function of the transverse coordinate $y$ across the ribbon width.
The results correspond to an average over $1000$ realizations, where we set $E=-0.001\ eV$, $L=20\Delta$, $W=20\Delta$, $\Delta= 5a_0$ and $a_0=0.142\ nm$.
In panel (a) we use $U_{C}=0$, $U_{Dis}=0$, $U_{SOC}=0.01\ eV$ and vary the SOC disorder concentration $n_{SOC}$.
In (b) we vary the SOC strenght $U_{SOC}$ by keeping the concentration constant $n_{SOC}=5\%$.
We turn on the SOC coulomb contribution $U_C$ on panels (c) and (d).
The SOC strenght is $U_{SOC}=0.01\ eV$ and the concentration is $n_{SOC}=5\%$.
In (c) the coulomb potential assumes positive values while assuming negative values in (d). 
}
\label{fig:SQHE-Adatoms}
\end{figure}

Figure~\ref{fig:Injvsx} shows a single realization of the spin-up injection density $\Delta n^\uparrow$ in units of $1/\Delta^2 V$ (CHECK) for a symmetric bias.
The disorder concentration is $n_c=5 \%$  and the SO disorder strength is $U_{SOC}=0.02\ eV$. 
The injection properties are calculated at electronic energy  $E=-0.001\ eV$.
We find a clear and strong spin-up imbalance $\Delta_{\pm}^\uparrow = \Delta n^\uparrow(y=0)-\Delta n^\uparrow(y=W) > 0$ between the opposite edges of the system.
Due to the symmetries of the SO interaction, the spin down injection density $\Delta n^\downarrow$ (not shown here) produces the same absolute imbalance but with opposite sign $\Delta_{\pm}^\downarrow = -\Delta_{\pm}^\uparrow$.
The results indicate that the QSHE is indeed present with opposite spin Hall voltages $V_{sH}^\downarrow = -V_{sH}^\uparrow$.

In order to study the dependence of the QSHE on the model parameters we plot the average over disorder configurations of the transverse section of the spin up injection density $\Delta n^\uparrow$ across the width in Fig.~\ref{fig:SQHE-Adatoms}.
First we turn off the local coulomb potential $U_C=0$ and the disorder potential $U_{Dis}=0$.
We vary the concentration $n_C$ in Fig.~\ref{fig:SQHE-Adatoms}(a) using a local spin-orbit coupling $U_{SOC}=0.01\ eV$.
In the absence of adatoms $n_c=0$ there is no injection imbalance ($\Delta_{\pm}^\uparrow=0$).
As we increase the concentration up to $n_c=8\%$ the injection imbalance $\Delta_{\pm}^\uparrow$ increases as well showing a monotonic increase in the interval $n_c\in [0,8\%]$.
Figure~\ref{fig:SQHE-Adatoms}(b) shows the dependence of $\Delta n^\uparrow$ on the local spin-orbit coupling $U_{SOC}$ for $U_C=0$ and $n_c=5\%$. 
We find that the injection imbalance also increases monotonically with the local spin-orbit strength $U_{SOC}$.

In Figs.~\ref{fig:SQHE-Adatoms}(c) and \ref{fig:SQHE-Adatoms}(d) we keep $U_{SOC}$ and $n_C$ constant, turning on the local coulomb potential to analyze the dependence of $\Delta n^\uparrow$ on $U_C$.
Figure~\ref{fig:SQHE-Adatoms}(c) shows that by increasing $U_C$, the injection imbalance $\Delta_{\pm}^\uparrow$ decreases for positive values of $U_C$.
The injection density $\Delta n^\uparrow(y=W)$ remains roughly constant while $\Delta n^\uparrow(y=0)$ decreases in the interval $[0.005,0.020]\ eV$, decreasing $\Delta_{\pm}^\uparrow$ as a consequence.
On the other hand, Fig.~\ref{fig:SQHE-Adatoms}(d) shows the opposite behavior for negative values of $U_C$.
As $U_C$ varies from $U_C=0$ to $U_C=-0.020\ eV$, the imbalance $\Delta_{\pm}^\uparrow$ increases.  
Thus, spin up Hall voltage $V_{sH}^\uparrow$ decreases with $|U_C|$ for adatoms with positive Coulomb interaction and increases with $|U_C|$ for adatoms with negative Coulomb interaction. 

\begin{figure}[!t]
\includegraphics[width=0.85\columnwidth]{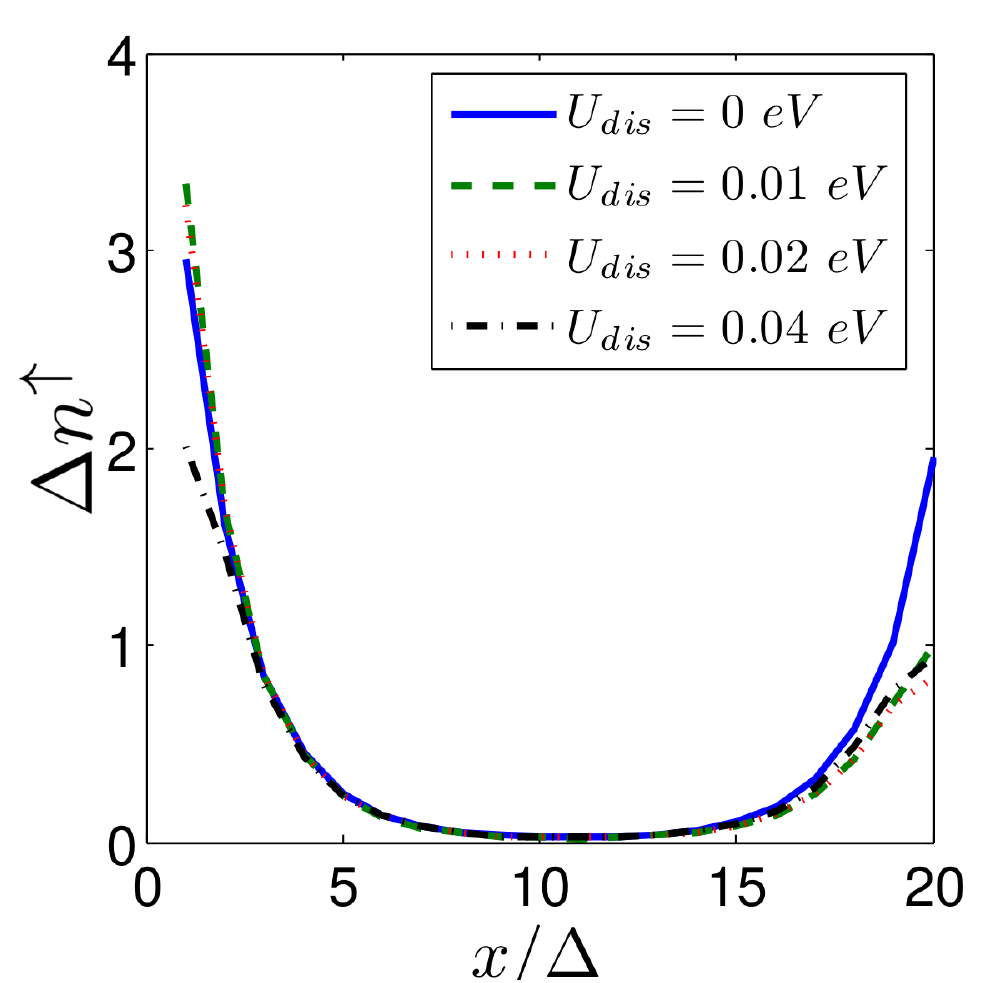}
\caption{
Spin-up injection density $\Delta n^\uparrow$ in units of $1/(\Delta^2 V)$ calculated at $x=L/2$ as a function of the transverse coordinate $y$ across the ribbon width.
The results correspond to an average over $1000$ realizations, where we set $E=-0.001\ eV$, $L=20\Delta$, $W=20\Delta$, $\Delta= 5a_0$ and $a_0=0.142\ nm$.
We use $U_{SOC}=0.01\ eV$, $U_C=0$, $n_{SOC}=5\%$, $n_{Dis}=5\%$ and vary the non-magnetic disorder strength $U_{Dis}$ from $0$ to $0.04\ eV$.
See the main text for more details.
}
\label{fig:003}
\end{figure}

Next we consider the diagonal disorder produced by a different source $U_{Dis}(\mathbf r)$.
These additional scatterers are nonmagnetic and are placed in positions that are different than the adatoms positions.
In this case the system has a coverage $n_{SOC}=5\%$ of adatoms with SOC and Coulomb strengths $U_{SOC}=0.01\ eV$ and $U_C=0$.
Figure~\ref{fig:003} shows the spin-up injection density $\Delta n^\uparrow$ as a function of the non-magnetic disorder strength $U_{Dis}$ for a non-magnetic disorder coverage $n_{Dis}=5\%$. 
The plot shows that $\Delta n^\uparrow$ does not vary monotonically with the disorder strength $U_{Dis}$.
The injection imbalance between the edges $\Delta_{\pm}^\uparrow$ increases when $U_{Dis}$ varies from $0$ to $0.01\ eV$ and decreases when $U_{Dis}$ varies from $0.01\ eV$ to $0.04\ eV$.
As a matter of fact, there is an optimal value of the disorder strength, which is rougly $U_{dis}=0.01\ eV$, that maximizes the injection imbalance $\Delta_{\pm}^\uparrow$ between opposite edges and the spin Hall voltages $V_{sH}^s$ as a consequence.

\subsection{Green's function approach applied to the quantum spin Hall effect on graphene}
\label{sec:greensfunctions}

In this section we present our numerical results on the spin-resolved injection density $\Delta n^s$ for a finite size graphene nanoribbon doped with adatoms using the tight-binding model proposed by Ref.~\onlinecite{Weeks11}.
Within this model, we also expect an imbalance $\Delta_{\pm}^s \neq 0$ in the spin-resolved injection density due to the QSHE generated by the presence of adatoms that act as spin-orbit scattering centers, for electronic energies near the charge neutrality point.  

The tight-binding description of graphene including the contribution from the adatoms reads \cite{Weeks11,Kane05,Shevtsov12prx,Jiang12}
\begin{align}
	H = -t\sum_{\left\langle i,j\right\rangle,\sigma} c_{i\sigma}^\dagger c_{j\sigma}
	+ \sum_{j,\sigma} \epsilon_j c_{j\sigma}^\dagger c_{j\sigma}\nonumber\\
	+ it_{\rm SO} \sum_{\left\langle\left\langle i,j\right\rangle\right\rangle,\sigma\sigma'} \nu_{ij} c_{i\sigma}^\dagger s^z_{\sigma\sigma'} c_{j\sigma'} .
	\label{H_QSH}
\end{align}
The operator $c_{i\sigma}^\dagger$ ($c_{j\sigma}$) creates (destroys) an electron with spin $\sigma=\uparrow,\downarrow$ at the site $i$ ($j$).
The hopping integral between first neighbors in the kinetic term has value $t=2.7$ eV, $\epsilon_j$ are on-site energies randomly chosen from a uniform distribution in the interval $[-V_{Dis},V_{Dis}]$, where $V_{Dis}$ is the disorder strength. 
The spin-orbit interaction due to adatoms in the third term has strength $t_{\rm SO}$ and acts only between second neighbors around the adatom, which is placed at the center of the corresponding hexagon. 
The Pauli matrix $s^z$ ensures that the hopping has opposite signs for different spin orientations while $\nu_{ij}$ distinguishes between the clockwise ($\nu_{ij}=1$) and the counterclockwise ($\nu_{ij}=-1$) directions \cite{Weeks11}. 

One of the most efficient ways to calculate transport properties of two terminal systems is the Green's functions technique \cite{Datta97,Haug08}. 
We calculate the non-equilibrium injectivity in Eq.~(\ref{eq:inj}) in terms of equilibrium Green's functions as \cite{Hernandez13}
\begin{align}
\frac{dn({\bf r},\alpha)}{dE} = \int_{-\infty}^\infty \frac{dE}{2\pi}
\left(-\frac{\partial f_0}{\partial E}\right) \langle {\bf r} | G^r_0
\Gamma_\alpha G^a_0 | {\bf r} \rangle,
\label{eq:NEGF-inj}
\end{align}
where $G^r_0$ ($G^a_0$) is the equilibrium retarded (advanced) Green's function and $\Gamma_\alpha$ is the linewidth function of the lead $\alpha$.
We calculate $G^r_0$ and $\Gamma_\alpha$ at a given electronic energy $E$ by means of the recursive Green's function technique (RGF) \cite{MacKinnon85,Lewenkopf13,Lima18} and decimation \cite{Sancho85,Lewenkopf13}, respectively. 
We compute the advanced Green's function via its standard relation with the retarded one, namely, $G^a_0=(G^r_0)^\dagger$. 
At zero temperature, the injectivity in Eq.~(\ref{eq:NEGF-inj}) yields
\begin{align}
	\frac{dn^s(\mathbf r,\alpha)}{dE} = \frac{1}{2\pi} \langle {\bf r} | G^{r,ss}_0(E_F)
	\Gamma_\alpha^{ss}(E_F) G^{a,ss}_0(E_F) | {\bf r} \rangle,
	\label{injRGF}
\end{align}
where $E_F$ is the Fermi energy at equilibrium, $G^{r,ss}_0$ ($G^{a,ss}_0$) is the equilibrium retarded (advanced) Green's function block connecting the same spin orientation $s=\uparrow,\downarrow$.
We assume that up and down spin components are equally injected in the system, $\Gamma_\alpha^{\downarrow\downarrow}=\Gamma_\alpha^{\uparrow\uparrow}$.
Our model system is a graphene nanoribbon with armchair edges along the transport direction with width and length equal to $100$\AA\ and $170$\AA, respectively. 
We attach two semi-infinite leads at $x=0$ (left) and $x=170$\AA\ (right).

\begin{figure}[!t]
	\includegraphics[width=0.75\columnwidth]{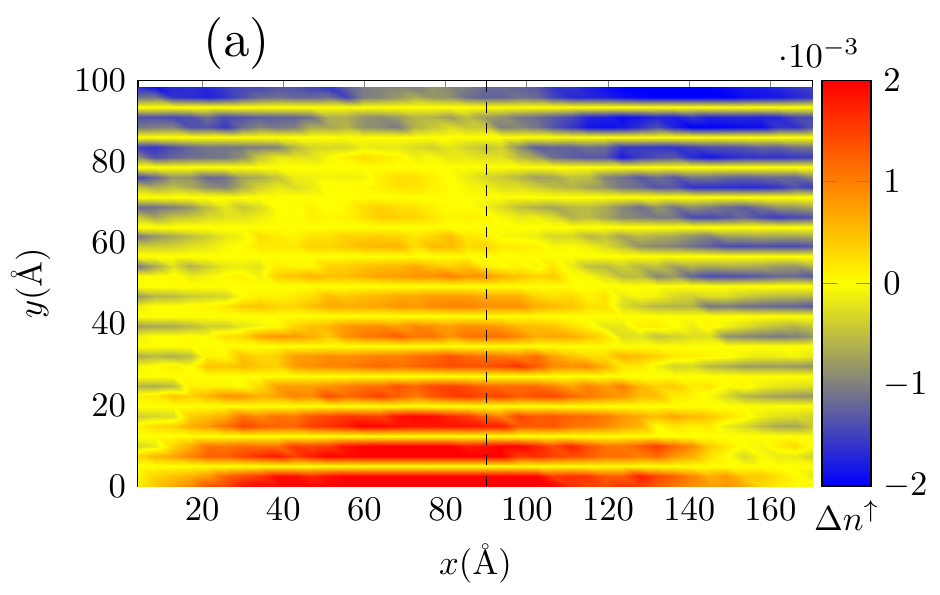}
	\includegraphics[width=0.75\columnwidth]{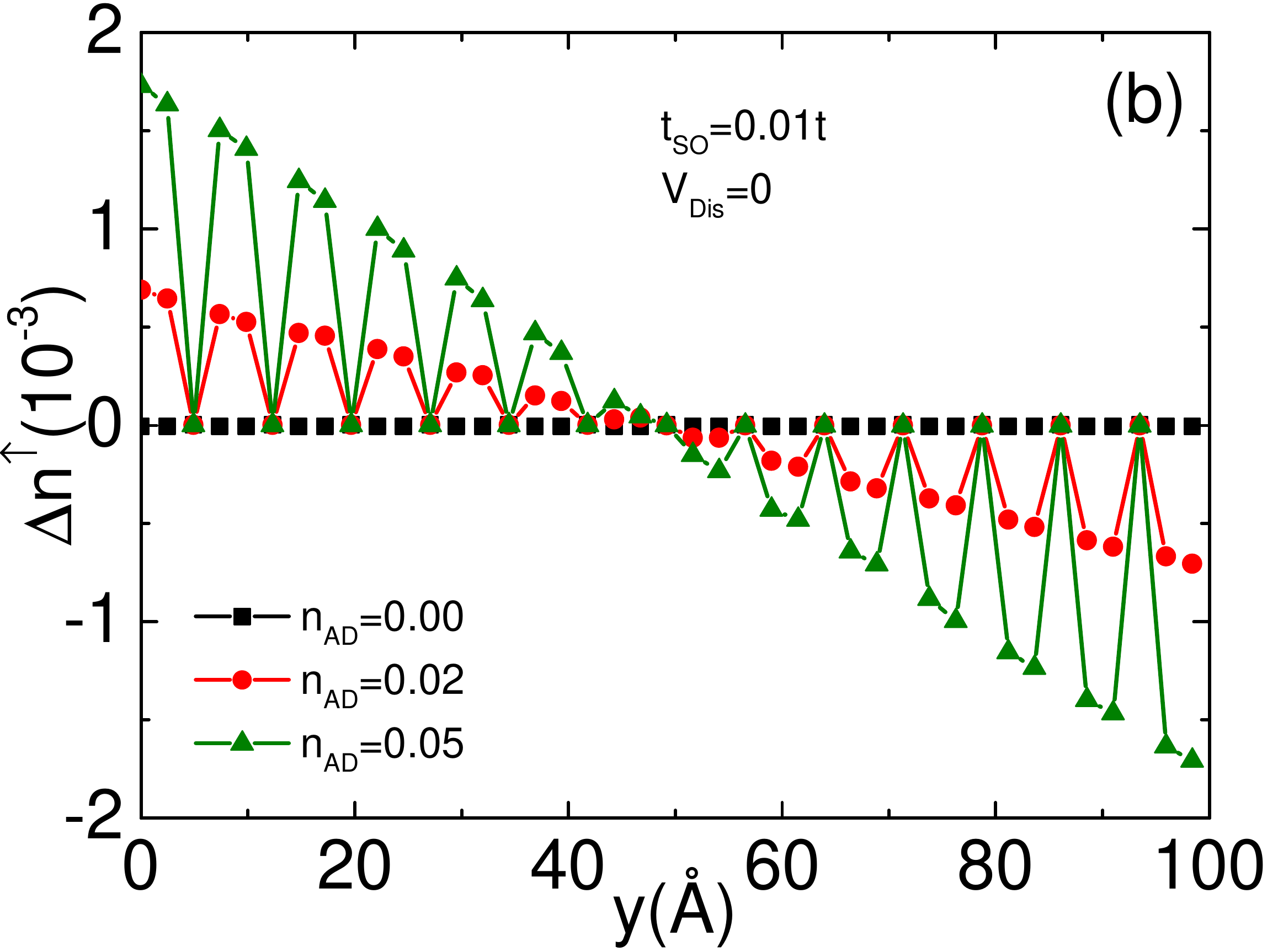}
	\includegraphics[width=0.75\columnwidth]{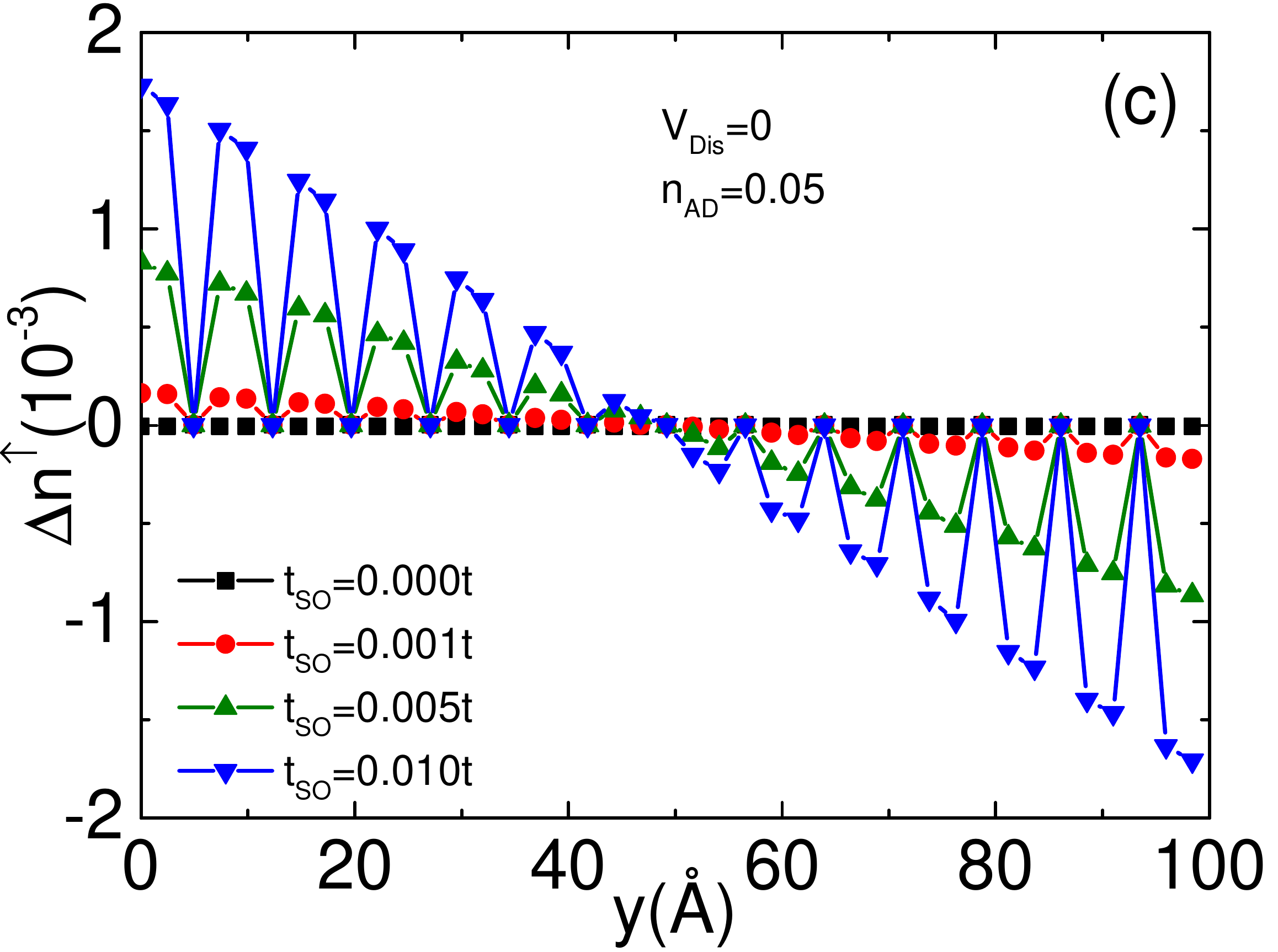}
	\includegraphics[width=0.75\columnwidth]{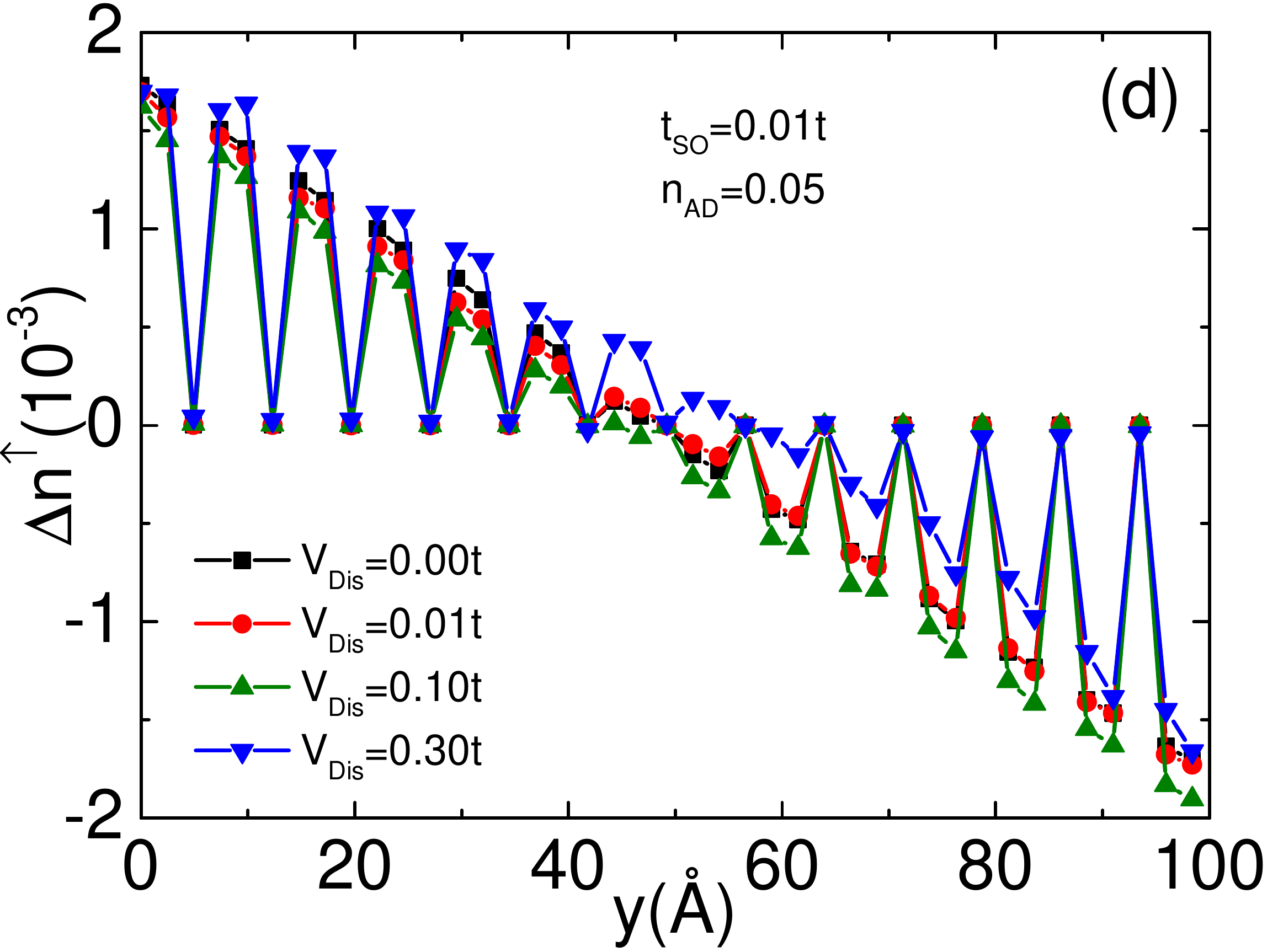}
\caption{
(Color online) (a) Spin-up injection density $\Delta n^\uparrow$ in units of $1/eV$ in a $100$\AA$\times$ $170$\AA\ graphene nanoribbon for a single realization, where $n_{AD}=0.05$, $t_{SO}=0.01t$ and $V_{Dis}=0$.
The dashed line indicates one of the cross sections used to calculate the average over all the cross sections that we show in panels (b)-(d) as a function of $y$.
The panels (b), (c) and (d) show the average value of $\Delta n^\uparrow$ for different values of adatom concentration $n_{AD}$, spin-orbit strength $t_{SO}$ and disorder strength $W$, respectively.
The electronic energy is $E_F= 0.01t$ in all results.
}
\label{fig:injrgf}
\end{figure}

Figure~\ref{fig:injrgf} shows the numerical results obtained by means of the spin-up component of Eq.~(\ref{injRGF}). 
We show the spin-up injection density $\Delta n^\uparrow$ for a single realization as the color map in Fig.~\ref{fig:injrgf}(a).
Analogously to the previous section, we find that $\Delta n^\uparrow$ is higher at one edge ($y=0$) than at the opposite one ($y=100$\AA).
The maximum and minimum values of $\Delta n^\uparrow$ at each cross section along the $y$ direction, indicated by the dashed line in Fig.~\ref{fig:injrgf}(a), are different due to the presence of the disordered distribution of adatoms in the system.

The panels (b), (c) and (d) of Fig.~\ref{fig:injrgf} show the average injection densities $\Delta n^\uparrow$ taken over all the cross sections in the system for different values of the adatom concentration $n_{AD}$, the spin-orbit strength $t_{SO}$ and the disorder strength $V_{Dis}$, respectively.
 From Fig.~\ref{fig:injrgf}(b) we find that $\Delta n^\uparrow$ vanishes in the absence of adatoms $n_{AD}=0$ (no spin-orbit) and increases at the edges as we increase the adatom concentration $n_{AD}$.
 The effect is most prominent at the edges and $\Delta n^\uparrow$ varies almost linearly from one edge to the opposite one.
 
We find a similar behavior in Fig.~\ref{fig:injrgf}(c) where $\Delta n^\uparrow$ vanishes in the absence of the spin-orbit strength ($t_{SO}=0$) and increases with its value having a approximately linear dependence with $y$.
In Fig.~\ref{fig:injrgf}(d) we show that these results are robust against disorder.
 As we increase the disorder strength $V_{Dis}$ we find only small fluctuations in $\Delta n^\uparrow$ compared to the case without diagonal non-magnetic disorder ($V_{Dis}=0$).
We find that the injection density imbalance $\Delta_{\pm}^\uparrow$ increase with both the adatom concentration and the spin-orbit strength $t_{SO}$, similarly to the previous section results.
On the other hand, the non-magnetic disorder within this model barely affects the imbalance $\Delta_{\pm}^\uparrow$, which is at odds with the previous section results. 
Therefore, the spin Hall voltage $V_{sH}^\uparrow$ increases with both the adatom concentration and the spin-orbit strength $t_{SO}$, and it is robust against non-magnetic disorder.
Theses results are in line with the conclusions drawn via full conductance calculations performed in Ref.~\cite{Weeks11}.

\section{Conclusions}
\label{sec:conclusions}

We presented a method to numerically study transverse conductances using a two-terminal setup.
We establish a connection between the transverse voltage in the system and the differences between the injectivities from each terminal.
The connection is derived using nonlinear transport concepts from the literature.

We applied our method  to study the QSHE in graphene doped with adatoms within two distinct models, a finite difference implementation of the Dirac Hamiltonian combined with the scattering matrix approach and a tight-binding Hamiltonian combined with the non-equilibrium Green's functions approach.
The results show that the presence of adatoms produces a considerable difference between the spin injection densities at the edges of the ribbon, resulting in a QSHE, i.e., leading to non-zero spin Hall voltages $V_{sH}^{\uparrow,\downarrow}$. 
The latter increase with the adatom concentration and the SOC strength for both models, which is inline with the expected behavior of the QSHE in graphene doped with adatoms for electronic energies near the charge neutrality point \cite{Kane05,Weeks11}.

In the model using the tight-binding description of graphene, we found that non-magnetic disorder does not affect the spin Hall voltages.
On the other hand, in the continuum description using the Dirac Hamiltonian, the non-magnetic disorder can increase or decrease the spin Hall voltages depending on both the intensity and the sign of the disorder strength. 
Furthermore, we find the optimal value of the non-magnetic disorder strength that maximizes the spin Hall voltages.
These results show that it is indeed possible to extract not only qualitative but also quantitative information of the system by studying the two-terminal injectivities using the proposed method.

\bibliography{refs2}

\begin{thebibliography}{29}%
\makeatletter
\providecommand \@ifxundefined [1]{%
 \@ifx{#1\undefined}
}%
\providecommand \@ifnum [1]{%
 \ifnum #1\expandafter \@firstoftwo
 \else \expandafter \@secondoftwo
 \fi
}%
\providecommand \@ifx [1]{%
 \ifx #1\expandafter \@firstoftwo
 \else \expandafter \@secondoftwo
 \fi
}%
\providecommand \natexlab [1]{#1}%
\providecommand \enquote  [1]{``#1''}%
\providecommand \bibnamefont  [1]{#1}%
\providecommand \bibfnamefont [1]{#1}%
\providecommand \citenamefont [1]{#1}%
\providecommand \href@noop [0]{\@secondoftwo}%
\providecommand \href [0]{\begingroup \@sanitize@url \@href}%
\providecommand \@href[1]{\@@startlink{#1}\@@href}%
\providecommand \@@href[1]{\endgroup#1\@@endlink}%
\providecommand \@sanitize@url [0]{\catcode `\\12\catcode `\$12\catcode
  `\&12\catcode `\#12\catcode `\^12\catcode `\_12\catcode `\%12\relax}%
\providecommand \@@startlink[1]{}%
\providecommand \@@endlink[0]{}%
\providecommand \url  [0]{\begingroup\@sanitize@url \@url }%
\providecommand \@url [1]{\endgroup\@href {#1}{\urlprefix }}%
\providecommand \urlprefix  [0]{URL }%
\providecommand \Eprint [0]{\href }%
\providecommand \doibase [0]{http://dx.doi.org/}%
\providecommand \selectlanguage [0]{\@gobble}%
\providecommand \bibinfo  [0]{\@secondoftwo}%
\providecommand \bibfield  [0]{\@secondoftwo}%
\providecommand \translation [1]{[#1]}%
\providecommand \BibitemOpen [0]{}%
\providecommand \bibitemStop [0]{}%
\providecommand \bibitemNoStop [0]{.\EOS\space}%
\providecommand \EOS [0]{\spacefactor3000\relax}%
\providecommand \BibitemShut  [1]{\csname bibitem#1\endcsname}%
\let\auto@bib@innerbib\@empty
\bibitem [{\citenamefont {Wang}\ and\ \citenamefont {Zhang}(2017)}]{Wang17}%
  \BibitemOpen
  \bibfield  {author} {\bibinfo {author} {\bibfnamefont {J.}~\bibnamefont
  {Wang}}\ and\ \bibinfo {author} {\bibfnamefont {S.-C.}\ \bibnamefont
  {Zhang}},\ }\href {http://dx.doi.org/10.1038/nmat5012} {\bibfield  {journal}
  {\bibinfo  {journal} {Nature Materials}\ }\textbf {\bibinfo {volume} {16}},\
  \bibinfo {pages} {1062 EP } (\bibinfo {year} {2017})},\ \bibinfo {note}
  {perspective}\BibitemShut {NoStop}%
\bibitem [{\citenamefont {Klitzing}\ \emph {et~al.}(1980)\citenamefont
  {Klitzing}, \citenamefont {Dorda},\ and\ \citenamefont
  {Pepper}}]{vonKlitzing80}%
  \BibitemOpen
  \bibfield  {author} {\bibinfo {author} {\bibfnamefont {K.~v.}\ \bibnamefont
  {Klitzing}}, \bibinfo {author} {\bibfnamefont {G.}~\bibnamefont {Dorda}}, \
  and\ \bibinfo {author} {\bibfnamefont {M.}~\bibnamefont {Pepper}},\ }\href
  {\doibase 10.1103/PhysRevLett.45.494} {\bibfield  {journal} {\bibinfo
  {journal} {Phys. Rev. Lett.}\ }\textbf {\bibinfo {volume} {45}},\ \bibinfo
  {pages} {494} (\bibinfo {year} {1980})}\BibitemShut {NoStop}%
\bibitem [{\citenamefont {Bernevig}\ \emph {et~al.}(2006)\citenamefont
  {Bernevig}, \citenamefont {Hughes},\ and\ \citenamefont
  {Zhang}}]{Bernevig06}%
  \BibitemOpen
  \bibfield  {author} {\bibinfo {author} {\bibfnamefont {B.~A.}\ \bibnamefont
  {Bernevig}}, \bibinfo {author} {\bibfnamefont {T.~L.}\ \bibnamefont
  {Hughes}}, \ and\ \bibinfo {author} {\bibfnamefont {S.-C.}\ \bibnamefont
  {Zhang}},\ }\href@noop {} {\bibfield  {journal} {\bibinfo  {journal}
  {Science}\ }\textbf {\bibinfo {volume} {314}},\ \bibinfo {pages} {1757}
  (\bibinfo {year} {2006})},\ \Eprint
  {http://arxiv.org/abs/http://www.sciencemag.org/content/314/5806/1757.full.pdf}
  {http://www.sciencemag.org/content/314/5806/1757.full.pdf} \BibitemShut
  {NoStop}%
\bibitem [{\citenamefont {K\"{o}nig}\ \emph {et~al.}(2007)\citenamefont
  {K\"{o}nig}, \citenamefont {Wiedmann}, \citenamefont {Br\"{u}ne},
  \citenamefont {Roth}, \citenamefont {Buhmann}, \citenamefont {Molenkamp},
  \citenamefont {Qi},\ and\ \citenamefont {Zhang}}]{Konig07}%
  \BibitemOpen
  \bibfield  {author} {\bibinfo {author} {\bibfnamefont {M.}~\bibnamefont
  {K\"{o}nig}}, \bibinfo {author} {\bibfnamefont {S.}~\bibnamefont {Wiedmann}},
  \bibinfo {author} {\bibfnamefont {C.}~\bibnamefont {Br\"{u}ne}}, \bibinfo
  {author} {\bibfnamefont {A.}~\bibnamefont {Roth}}, \bibinfo {author}
  {\bibfnamefont {H.}~\bibnamefont {Buhmann}}, \bibinfo {author} {\bibfnamefont
  {L.~W.}\ \bibnamefont {Molenkamp}}, \bibinfo {author} {\bibfnamefont {X.-L.}\
  \bibnamefont {Qi}}, \ and\ \bibinfo {author} {\bibfnamefont {S.-C.}\
  \bibnamefont {Zhang}},\ }\href {\doibase 10.1126/science.1148047} {\bibfield
  {journal} {\bibinfo  {journal} {Science}\ }\textbf {\bibinfo {volume}
  {318}},\ \bibinfo {pages} {766} (\bibinfo {year} {2007})}\BibitemShut
  {NoStop}%
\bibitem [{\citenamefont {Maciejko}\ \emph {et~al.}(2010)\citenamefont
  {Maciejko}, \citenamefont {Qi},\ and\ \citenamefont {Zhang}}]{Maciejko10}%
  \BibitemOpen
  \bibfield  {author} {\bibinfo {author} {\bibfnamefont {J.}~\bibnamefont
  {Maciejko}}, \bibinfo {author} {\bibfnamefont {X.-L.}\ \bibnamefont {Qi}}, \
  and\ \bibinfo {author} {\bibfnamefont {S.-C.}\ \bibnamefont {Zhang}},\ }\href
  {\doibase 10.1103/PhysRevB.82.155310} {\bibfield  {journal} {\bibinfo
  {journal} {Phys. Rev. B}\ }\textbf {\bibinfo {volume} {82}},\ \bibinfo
  {pages} {155310} (\bibinfo {year} {2010})}\BibitemShut {NoStop}%
\bibitem [{\citenamefont {Scharf}\ \emph {et~al.}(2012)\citenamefont {Scharf},
  \citenamefont {Matos-Abiague},\ and\ \citenamefont {Fabian}}]{Scharf12}%
  \BibitemOpen
  \bibfield  {author} {\bibinfo {author} {\bibfnamefont {B.}~\bibnamefont
  {Scharf}}, \bibinfo {author} {\bibfnamefont {A.}~\bibnamefont
  {Matos-Abiague}}, \ and\ \bibinfo {author} {\bibfnamefont {J.}~\bibnamefont
  {Fabian}},\ }\href {\doibase 10.1103/PhysRevB.86.075418} {\bibfield
  {journal} {\bibinfo  {journal} {Phys. Rev. B}\ }\textbf {\bibinfo {volume}
  {86}},\ \bibinfo {pages} {075418} (\bibinfo {year} {2012})}\BibitemShut
  {NoStop}%
\bibitem [{\citenamefont {Durnev}\ and\ \citenamefont
  {Tarasenko}(2016)}]{Durnev16}%
  \BibitemOpen
  \bibfield  {author} {\bibinfo {author} {\bibfnamefont {M.~V.}\ \bibnamefont
  {Durnev}}\ and\ \bibinfo {author} {\bibfnamefont {S.~A.}\ \bibnamefont
  {Tarasenko}},\ }\href {\doibase 10.1103/PhysRevB.93.075434} {\bibfield
  {journal} {\bibinfo  {journal} {Phys. Rev. B}\ }\textbf {\bibinfo {volume}
  {93}},\ \bibinfo {pages} {075434} (\bibinfo {year} {2016})}\BibitemShut
  {NoStop}%
\bibitem [{\citenamefont {Nanclares}\ \emph {et~al.}(2017)\citenamefont
  {Nanclares}, \citenamefont {Lima}, \citenamefont {Lewenkopf},\ and\
  \citenamefont {da~Silva}}]{Nanclares17}%
  \BibitemOpen
  \bibfield  {author} {\bibinfo {author} {\bibfnamefont {D.}~\bibnamefont
  {Nanclares}}, \bibinfo {author} {\bibfnamefont {L.~R.~F.}\ \bibnamefont
  {Lima}}, \bibinfo {author} {\bibfnamefont {C.~H.}\ \bibnamefont {Lewenkopf}},
  \ and\ \bibinfo {author} {\bibfnamefont {L.~G. G. V.~D.}\ \bibnamefont
  {da~Silva}},\ }\href {\doibase 10.1103/PhysRevB.96.155302} {\bibfield
  {journal} {\bibinfo  {journal} {Phys. Rev. B}\ }\textbf {\bibinfo {volume}
  {96}},\ \bibinfo {pages} {155302} (\bibinfo {year} {2017})}\BibitemShut
  {NoStop}%
\bibitem [{\citenamefont {MacKinnon}(1985)}]{MacKinnon85}%
  \BibitemOpen
  \bibfield  {author} {\bibinfo {author} {\bibfnamefont {A.}~\bibnamefont
  {MacKinnon}},\ }\href {\doibase 10.1007/BF01328846} {\bibfield  {journal}
  {\bibinfo  {journal} {Zeitschrift f{\"u}r Physik B Condensed Matter}\
  }\textbf {\bibinfo {volume} {59}},\ \bibinfo {pages} {385} (\bibinfo {year}
  {1985})}\BibitemShut {NoStop}%
\bibitem [{\citenamefont {Lewenkopf}\ and\ \citenamefont
  {Mucciolo}(2013)}]{Lewenkopf13}%
  \BibitemOpen
  \bibfield  {author} {\bibinfo {author} {\bibfnamefont {C.}~\bibnamefont
  {Lewenkopf}}\ and\ \bibinfo {author} {\bibfnamefont {E.}~\bibnamefont
  {Mucciolo}},\ }\href {\doibase 10.1007/s10825-013-0458-7} {\bibfield
  {journal} {\bibinfo  {journal} {Journal of Computational Electronics}\
  }\textbf {\bibinfo {volume} {12}},\ \bibinfo {pages} {203} (\bibinfo {year}
  {2013})}\BibitemShut {NoStop}%
\bibitem [{\citenamefont {Baranger}\ \emph {et~al.}(1988)\citenamefont
  {Baranger}, \citenamefont {Stone},\ and\ \citenamefont
  {DiVincenzo}}]{Baranger88}%
  \BibitemOpen
  \bibfield  {author} {\bibinfo {author} {\bibfnamefont {H.~U.}\ \bibnamefont
  {Baranger}}, \bibinfo {author} {\bibfnamefont {A.~D.}\ \bibnamefont {Stone}},
  \ and\ \bibinfo {author} {\bibfnamefont {D.~P.}\ \bibnamefont {DiVincenzo}},\
  }\href {\doibase 10.1103/PhysRevB.37.6521} {\bibfield  {journal} {\bibinfo
  {journal} {Phys. Rev. B}\ }\textbf {\bibinfo {volume} {37}},\ \bibinfo
  {pages} {6521} (\bibinfo {year} {1988})}\BibitemShut {NoStop}%
\bibitem [{\citenamefont {Kazymyrenko}\ and\ \citenamefont
  {Waintal}(2008)}]{Kazymyrenko08}%
  \BibitemOpen
  \bibfield  {author} {\bibinfo {author} {\bibfnamefont {K.}~\bibnamefont
  {Kazymyrenko}}\ and\ \bibinfo {author} {\bibfnamefont {X.}~\bibnamefont
  {Waintal}},\ }\href {\doibase 10.1103/PhysRevB.77.115119} {\bibfield
  {journal} {\bibinfo  {journal} {Phys. Rev. B}\ }\textbf {\bibinfo {volume}
  {77}},\ \bibinfo {pages} {115119} (\bibinfo {year} {2008})}\BibitemShut
  {NoStop}%
\bibitem [{\citenamefont {Thorgilsson}\ \emph {et~al.}(2014)\citenamefont
  {Thorgilsson}, \citenamefont {Viktorsson},\ and\ \citenamefont
  {Erlingsson}}]{Thorgilsson14}%
  \BibitemOpen
  \bibfield  {author} {\bibinfo {author} {\bibfnamefont {G.}~\bibnamefont
  {Thorgilsson}}, \bibinfo {author} {\bibfnamefont {G.}~\bibnamefont
  {Viktorsson}}, \ and\ \bibinfo {author} {\bibfnamefont {S.}~\bibnamefont
  {Erlingsson}},\ }\href {\doibase 10.1016/j.jcp.2013.12.054} {\bibfield
  {journal} {\bibinfo  {journal} {J. Comp. Phys.}\ }\textbf {\bibinfo {volume}
  {261}},\ \bibinfo {pages} {256 } (\bibinfo {year} {2014})}\BibitemShut
  {NoStop}%
\bibitem [{\citenamefont {Lima}\ \emph {et~al.}(2018)\citenamefont {Lima},
  \citenamefont {Dusko},\ and\ \citenamefont {Lewenkopf}}]{Lima18}%
  \BibitemOpen
  \bibfield  {author} {\bibinfo {author} {\bibfnamefont {L.~R.~F.}\
  \bibnamefont {Lima}}, \bibinfo {author} {\bibfnamefont {A.}~\bibnamefont
  {Dusko}}, \ and\ \bibinfo {author} {\bibfnamefont {C.}~\bibnamefont
  {Lewenkopf}},\ }\href@noop {} {\enquote {\bibinfo {title} {An efficient
  method for computing the electronic transport properties of a multi-terminal
  system},}\ } (\bibinfo {year} {2018}),\ \Eprint
  {http://arxiv.org/abs/arXiv:1801.07298} {arXiv:1801.07298} \BibitemShut
  {NoStop}%
\bibitem [{\citenamefont {Weeks}\ \emph {et~al.}(2011)\citenamefont {Weeks},
  \citenamefont {Hu}, \citenamefont {Alicea}, \citenamefont {Franz},\ and\
  \citenamefont {Wu}}]{Weeks11}%
  \BibitemOpen
  \bibfield  {author} {\bibinfo {author} {\bibfnamefont {C.}~\bibnamefont
  {Weeks}}, \bibinfo {author} {\bibfnamefont {J.}~\bibnamefont {Hu}}, \bibinfo
  {author} {\bibfnamefont {J.}~\bibnamefont {Alicea}}, \bibinfo {author}
  {\bibfnamefont {M.}~\bibnamefont {Franz}}, \ and\ \bibinfo {author}
  {\bibfnamefont {R.}~\bibnamefont {Wu}},\ }\href {\doibase
  10.1103/PhysRevX.1.021001} {\bibfield  {journal} {\bibinfo  {journal} {Phys.
  Rev. X}\ }\textbf {\bibinfo {volume} {1}},\ \bibinfo {pages} {021001}
  (\bibinfo {year} {2011})}\BibitemShut {NoStop}%
\bibitem [{\citenamefont {Balakrishnan}\ \emph {et~al.}(2014)\citenamefont
  {Balakrishnan}, \citenamefont {Koon}, \citenamefont {Avsar}, \citenamefont
  {Ho}, \citenamefont {Lee}, \citenamefont {Jaiswal}, \citenamefont {Baeck},
  \citenamefont {Ahn}, \citenamefont {Ferreira}, \citenamefont {Cazalilla},
  \citenamefont {Castro~Neto},\ and\ \citenamefont
  {\"{O}zyilmaz}}]{Balakrishnan14}%
  \BibitemOpen
  \bibfield  {author} {\bibinfo {author} {\bibfnamefont {J.}~\bibnamefont
  {Balakrishnan}}, \bibinfo {author} {\bibfnamefont {G.}~\bibnamefont {Koon}},
  \bibinfo {author} {\bibfnamefont {A.}~\bibnamefont {Avsar}}, \bibinfo
  {author} {\bibfnamefont {Y.}~\bibnamefont {Ho}}, \bibinfo {author}
  {\bibfnamefont {J.}~\bibnamefont {Lee}}, \bibinfo {author} {\bibfnamefont
  {M.}~\bibnamefont {Jaiswal}}, \bibinfo {author} {\bibfnamefont {S.-J.}\
  \bibnamefont {Baeck}}, \bibinfo {author} {\bibfnamefont {J.-H.}\ \bibnamefont
  {Ahn}}, \bibinfo {author} {\bibfnamefont {A.}~\bibnamefont {Ferreira}},
  \bibinfo {author} {\bibfnamefont {M.}~\bibnamefont {Cazalilla}}, \bibinfo
  {author} {\bibfnamefont {A.}~\bibnamefont {Castro~Neto}}, \ and\ \bibinfo
  {author} {\bibfnamefont {B.}~\bibnamefont {\"{O}zyilmaz}},\ }\href {\doibase
  10.1038/ncomms5748} {\bibfield  {journal} {\bibinfo  {journal} {Nat.
  Commun.}\ }\textbf {\bibinfo {volume} {5}},\ \bibinfo {pages} {4748}
  (\bibinfo {year} {2014})}\BibitemShut {NoStop}%
\bibitem [{\citenamefont {Hern\'andez}\ and\ \citenamefont
  {Lewenkopf}(2012)}]{DFGAlexis12}%
  \BibitemOpen
  \bibfield  {author} {\bibinfo {author} {\bibfnamefont {A.~R.}\ \bibnamefont
  {Hern\'andez}}\ and\ \bibinfo {author} {\bibfnamefont {C.~H.}\ \bibnamefont
  {Lewenkopf}},\ }\href {\doibase 10.1103/PhysRevB.86.155439} {\bibfield
  {journal} {\bibinfo  {journal} {Phys. Rev. B}\ }\textbf {\bibinfo {volume}
  {86}},\ \bibinfo {pages} {155439} (\bibinfo {year} {2012})}\BibitemShut
  {NoStop}%
\bibitem [{\citenamefont {Castro~Neto}\ and\ \citenamefont
  {Guinea}(2009{\natexlab{a}})}]{CastroNeto09}%
  \BibitemOpen
  \bibfield  {author} {\bibinfo {author} {\bibfnamefont {A.~H.}\ \bibnamefont
  {Castro~Neto}}\ and\ \bibinfo {author} {\bibfnamefont {F.}~\bibnamefont
  {Guinea}},\ }\href {\doibase 10.1103/PhysRevLett.103.026804} {\bibfield
  {journal} {\bibinfo  {journal} {Phys. Rev. Lett.}\ }\textbf {\bibinfo
  {volume} {103}},\ \bibinfo {pages} {026804} (\bibinfo {year}
  {2009}{\natexlab{a}})}\BibitemShut {NoStop}%
\bibitem [{\citenamefont {{Hern\'andez, Alexis R.}}\ and\ \citenamefont
  {{Lewenkopf, Caio H.}}(2013)}]{Hernandez13}%
  \BibitemOpen
  \bibfield  {author} {\bibinfo {author} {\bibnamefont {{Hern\'andez, Alexis
  R.}}}\ and\ \bibinfo {author} {\bibnamefont {{Lewenkopf, Caio H.}}},\ }\href
  {\doibase 10.1140/epjb/e2013-31089-1} {\bibfield  {journal} {\bibinfo
  {journal} {Eur. Phys. J. B}\ }\textbf {\bibinfo {volume} {86}},\ \bibinfo
  {pages} {131} (\bibinfo {year} {2013})}\BibitemShut {NoStop}%
\bibitem [{\citenamefont {Landauer}(1987)}]{Landauer87}%
  \BibitemOpen
  \bibfield  {author} {\bibinfo {author} {\bibfnamefont {R.}~\bibnamefont
  {Landauer}},\ }\enquote {\bibinfo {title} {Nonlinearity: Historical and
  technological view},}\ in\ \href {\doibase 10.1007/978-3-642-83033-4_1}
  {\emph {\bibinfo {booktitle} {Nonlinearity in Condensed Matter: Proceedings
  of the Sixth Annual Conference, Center for Nonlinear Studies, Los Alamos, New
  Mexico, 5--9 May, 1986}}},\ \bibinfo {editor} {edited by\ \bibinfo {editor}
  {\bibfnamefont {A.~R.}\ \bibnamefont {Bishop}}, \bibinfo {editor}
  {\bibfnamefont {D.~K.}\ \bibnamefont {Campbell}}, \bibinfo {editor}
  {\bibfnamefont {P.}~\bibnamefont {Kumar}}, \ and\ \bibinfo {editor}
  {\bibfnamefont {S.~E.}\ \bibnamefont {Trullinger}}}\ (\bibinfo  {publisher}
  {Springer Berlin Heidelberg},\ \bibinfo {address} {Berlin, Heidelberg},\
  \bibinfo {year} {1987})\ pp.\ \bibinfo {pages} {2--22}\BibitemShut {NoStop}%
\bibitem [{\citenamefont {Buttiker}(1993)}]{Buttiker93}%
  \BibitemOpen
  \bibfield  {author} {\bibinfo {author} {\bibfnamefont {M.}~\bibnamefont
  {Buttiker}},\ }\href {http://stacks.iop.org/0953-8984/5/i=50/a=017}
  {\bibfield  {journal} {\bibinfo  {journal} {Journal of Physics: Condensed
  Matter}\ }\textbf {\bibinfo {volume} {5}},\ \bibinfo {pages} {9361} (\bibinfo
  {year} {1993})}\BibitemShut {NoStop}%
\bibitem [{\citenamefont {Bruus}\ and\ \citenamefont
  {Flensberg}(2004)}]{Bruus-Flensberg04}%
  \BibitemOpen
  \bibfield  {author} {\bibinfo {author} {\bibfnamefont {H.}~\bibnamefont
  {Bruus}}\ and\ \bibinfo {author} {\bibfnamefont {K.}~\bibnamefont
  {Flensberg}},\ }\href@noop {} {\emph {\bibinfo {title} {Many-body quantum
  theory in condensed matter physics - an introduction}}}\ (\bibinfo
  {publisher} {Oxford University Press},\ \bibinfo {year} {2004})\BibitemShut
  {NoStop}%
\bibitem [{\citenamefont {Castro~Neto}\ and\ \citenamefont
  {Guinea}(2009{\natexlab{b}})}]{AdatomsSOC}%
  \BibitemOpen
  \bibfield  {author} {\bibinfo {author} {\bibfnamefont {A.~H.}\ \bibnamefont
  {Castro~Neto}}\ and\ \bibinfo {author} {\bibfnamefont {F.}~\bibnamefont
  {Guinea}},\ }\href {\doibase 10.1103/PhysRevLett.103.026804} {\bibfield
  {journal} {\bibinfo  {journal} {Phys. Rev. Lett.}\ }\textbf {\bibinfo
  {volume} {103}},\ \bibinfo {pages} {026804} (\bibinfo {year}
  {2009}{\natexlab{b}})}\BibitemShut {NoStop}%
\bibitem [{\citenamefont {Kane}\ and\ \citenamefont {Mele}(2005)}]{Kane05}%
  \BibitemOpen
  \bibfield  {author} {\bibinfo {author} {\bibfnamefont {C.~L.}\ \bibnamefont
  {Kane}}\ and\ \bibinfo {author} {\bibfnamefont {E.~J.}\ \bibnamefont
  {Mele}},\ }\href {\doibase 10.1103/PhysRevLett.95.226801} {\bibfield
  {journal} {\bibinfo  {journal} {Phys. Rev. Lett.}\ }\textbf {\bibinfo
  {volume} {95}},\ \bibinfo {pages} {226801} (\bibinfo {year}
  {2005})}\BibitemShut {NoStop}%
\bibitem [{\citenamefont {Shevtsov}\ \emph {et~al.}(2012)\citenamefont
  {Shevtsov}, \citenamefont {Carmier}, \citenamefont {Petitjean}, \citenamefont
  {Groth}, \citenamefont {Carpentier},\ and\ \citenamefont
  {Waintal}}]{Shevtsov12prx}%
  \BibitemOpen
  \bibfield  {author} {\bibinfo {author} {\bibfnamefont {O.}~\bibnamefont
  {Shevtsov}}, \bibinfo {author} {\bibfnamefont {P.}~\bibnamefont {Carmier}},
  \bibinfo {author} {\bibfnamefont {C.}~\bibnamefont {Petitjean}}, \bibinfo
  {author} {\bibfnamefont {C.}~\bibnamefont {Groth}}, \bibinfo {author}
  {\bibfnamefont {D.}~\bibnamefont {Carpentier}}, \ and\ \bibinfo {author}
  {\bibfnamefont {X.}~\bibnamefont {Waintal}},\ }\href {\doibase
  10.1103/PhysRevX.2.031004} {\bibfield  {journal} {\bibinfo  {journal} {Phys.
  Rev. X}\ }\textbf {\bibinfo {volume} {2}},\ \bibinfo {pages} {031004}
  (\bibinfo {year} {2012})}\BibitemShut {NoStop}%
\bibitem [{\citenamefont {Jiang}\ \emph {et~al.}(2012)\citenamefont {Jiang},
  \citenamefont {Qiao}, \citenamefont {Liu}, \citenamefont {Shi},\ and\
  \citenamefont {Niu}}]{Jiang12}%
  \BibitemOpen
  \bibfield  {author} {\bibinfo {author} {\bibfnamefont {H.}~\bibnamefont
  {Jiang}}, \bibinfo {author} {\bibfnamefont {Z.}~\bibnamefont {Qiao}},
  \bibinfo {author} {\bibfnamefont {H.}~\bibnamefont {Liu}}, \bibinfo {author}
  {\bibfnamefont {J.}~\bibnamefont {Shi}}, \ and\ \bibinfo {author}
  {\bibfnamefont {Q.}~\bibnamefont {Niu}},\ }\href {\doibase
  10.1103/PhysRevLett.109.116803} {\bibfield  {journal} {\bibinfo  {journal}
  {Phys. Rev. Lett.}\ }\textbf {\bibinfo {volume} {109}},\ \bibinfo {pages}
  {116803} (\bibinfo {year} {2012})}\BibitemShut {NoStop}%
\bibitem [{\citenamefont {Datta}(1997)}]{Datta97}%
  \BibitemOpen
  \bibfield  {author} {\bibinfo {author} {\bibfnamefont {S.}~\bibnamefont
  {Datta}},\ }\href
  {http://www.amazon.com/exec/obidos/redirect?tag=citeulike07-20\&path=ASIN/0521599431}
  {\emph {\bibinfo {title} {{Electronic Transport in Mesoscopic Systems
  (Cambridge Studies in Semiconductor Physics and Microelectronic
  Engineering)}}}}\ (\bibinfo  {publisher} {Cambridge University Press},\
  \bibinfo {year} {1997})\BibitemShut {NoStop}%
\bibitem [{\citenamefont {Haug}\ and\ \citenamefont {Jauho}(2008)}]{Haug08}%
  \BibitemOpen
  \bibfield  {author} {\bibinfo {author} {\bibfnamefont {H.}~\bibnamefont
  {Haug}}\ and\ \bibinfo {author} {\bibfnamefont {A.~J.}\ \bibnamefont
  {Jauho}},\ }\href {\doibase 10.1007/978-3-540-73564-9} {\emph {\bibinfo
  {title} {Quantum Kinetics in Transport and Optics of Semiconductors}}},\
  \bibinfo {series} {Solid-State Sciences}, Vol.\ \bibinfo {volume} {123}\
  (\bibinfo  {publisher} {Springer Berlin Heidelberg},\ \bibinfo {address}
  {Berlin, Heidelberg},\ \bibinfo {year} {2008})\BibitemShut {NoStop}%
\bibitem [{\citenamefont {Sancho}\ \emph {et~al.}(1985)\citenamefont {Sancho},
  \citenamefont {Sancho}, \citenamefont {Sancho},\ and\ \citenamefont
  {Rubio}}]{Sancho85}%
  \BibitemOpen
  \bibfield  {author} {\bibinfo {author} {\bibfnamefont {M.~P.~L.}\
  \bibnamefont {Sancho}}, \bibinfo {author} {\bibfnamefont {J.~M.~L.}\
  \bibnamefont {Sancho}}, \bibinfo {author} {\bibfnamefont {J.~M.~L.}\
  \bibnamefont {Sancho}}, \ and\ \bibinfo {author} {\bibfnamefont
  {J.}~\bibnamefont {Rubio}},\ }\href
  {http://stacks.iop.org/0305-4608/15/i=4/a=009} {\bibfield  {journal}
  {\bibinfo  {journal} {Journal of Physics F: Metal Physics}\ }\textbf
  {\bibinfo {volume} {15}},\ \bibinfo {pages} {851} (\bibinfo {year}
  {1985})}\BibitemShut {NoStop}%
\end{thebibliography}%

\end{document}